\documentclass[reprint,
superscriptaddress,
 amsmath,amssymb,
 aps,
prb,
]{revtex4-1}

\usepackage{graphicx}
\usepackage{dcolumn}
\usepackage{bm}

\newcommand{\ve}[1]{\boldsymbol{#1}} 

\begin{document}

\preprint{APS/123-QED}

\title{Benchmarking theoretical electronic structure methods with photoemission orbital tomography}

\author{Anja Haags}
\author{Xiaosheng Yang}
\affiliation{Peter Gr\"unberg Institut (PGI-3), Forschungszentrum J\"ulich, 52425 J\"ulich, Germany}
\affiliation{J\"ulich Aachen Research Alliance (JARA), Fundamentals of Future Information Technology, 52425 J\"ulich, Germany}
\affiliation{Experimentalphysik IV A, RWTH Aachen University; Aachen, Germany}

\author{Larissa Egger}
\author{Dominik Brandstetter}
\affiliation{Institute of Physics, NAWI Graz, University of Graz, 8010 Graz, Austria}%

\author{Hans Kirschner}
\affiliation{Physikalisch-Technische Bundesanstalt (PTB), 10587 Berlin, Germany.}

\author{Alexander Gottwald}
\author{Mathias Richter}
\affiliation{Physikalisch-Technische Bundesanstalt (PTB), 10587 Berlin, Germany.}

\author{Georg Koller}
\author{Michael G. Ramsey}
\affiliation{Institute of Physics, NAWI Graz, University of Graz, 8010 Graz, Austria}%

\author{Fran\c{c}ois C. Bocquet}
\author{Serguei Soubatch}
\affiliation{Peter Gr\"unberg Institut (PGI-3), Forschungszentrum J\"ulich, 52425 J\"ulich, Germany}
\affiliation{J\"ulich Aachen Research Alliance (JARA), Fundamentals of Future Information Technology, 52425 J\"ulich, Germany}

\author{F. Stefan Tautz}
\affiliation{Peter Gr\"unberg Institut (PGI-3), Forschungszentrum J\"ulich, 52425 J\"ulich, Germany}
\affiliation{J\"ulich Aachen Research Alliance (JARA), Fundamentals of Future Information Technology, 52425 J\"ulich, Germany}
\affiliation{Experimentalphysik IV A, RWTH Aachen University; Aachen, Germany}

\author{Peter Puschnig}%
 \email{peter.puschnig@uni-graz.at}
\affiliation{Institute of Physics, NAWI Graz, University of Graz, 8010 Graz, Austria}%

\date{\today}

\begin{abstract}
In the past decade, photoemission orbital tomography (POT) has evolved into a powerful tool to investigate the electronic structure of organic molecules adsorbed on surfaces. By measuring the angular distribution of photoelectrons as a function of binding energy and making use of the momentum-space signature of molecular orbitals, POT leads to an orbital-resolved picture of the electronic density of states at the organic/metal interface. In this combined experimental and theoretical work, we apply POT to the prototypical organic $\pi$-conjugated molecule bisanthene (C$_{28}$H$_{14}$) which forms a highly oriented monolayer on a Cu(110) surface. Experimentally, we identify an unprecedented number of 13 $\pi$ and 12 $\sigma$ orbitals of bisanthene and measure their respective binding energies and spectral lineshapes at the bisanthene/Cu(110) interface. Theoretically, we perform density functional calculations for this interface employing four widely used exchange-correlation functionals from the families of the generalized gradient approximations as well as global and range-separated hybrid functionals. By analyzing the electronic structure in terms of orbital-projected density of states, we arrive at a detailed orbital-by-orbital assessment of theory vs. experiment. This allows us to benchmark the performance of the investigated functionals with regards to their capability of accounting for the orbital energy alignment at organic/metal interfaces.

\end{abstract}

\maketitle


\section{\label{sec:intro}Introduction}

Interfaces between organic molecules and metals play a central role in functional devices in nanoscience and nanotechnology.\cite{Ishii1999,Kahn2003,Ueno2008,Braun2009,MoleculeMetalInterface,Willenbockel2014,Liu2017,Ferri2017} 
When a molecule adsorbs on a metallic surface, several processes determine the resulting electronic structure of the interface.
First, the molecule gets physically attracted by van der Waals interactions.\cite{Berland2015,Hermann2017} 
Then, the ionization potential and electron affinity levels of the molecule are renormalized because of the proximity to the metal or, in a dense layer, also because of neighboring molecules as a result of
polarization effects.\cite{Neaton2006,Garcia-Lastra2009,Thygesen2009,Soubatch2009,Puschnig2012}
When the molecule approaches the substrate closer, 
the molecular orbitals start to overlap with the metallic states, and thus hybridized states
arise at the interface.\cite{Yamane2007,Ziroff2010,Berkebile2011,Wiessner2012b,Ules2014,Yang2022}
If the resulting level alignment permits, the native electronic states may be further reconfigured and charge
may be transferred, resulting in the population of formerly unoccupied molecular states or the depopulation of formerly occupied ones. Many examples of such charge transfers are known.\cite{Duhm2008,Rangger2009,Puschnig2009a,Ziroff2010,Heimel2013,Hofmann2015,Schonauer2016,Hollerer2017}

Ideally, all these mechanisms should be taken into account when predicting the electronic level alignment at organic/metal interfaces. The most commonly applied framework for this purpose is density functional theory (DFT) which has, however, several well-known limitations. First, it remains an ongoing challenge to correctly account for van der Waals dispersion interactions, which are often considered to be only a small contribution to the total energy, but are in fact also relevant for the electronic properties through the interplay of adsorption structure and electronic properties.\cite{Berland2015,Hermann2017}
Second, the common practice to interpret Kohn-Sham orbital energies as excitation energies is lacking rigorous theoretical justification. In the framework of generalized Kohn-Sham theory,\cite{Seidl1996,Kronik2012} it is indeed possible to identify the highest occupied Kohn-Sham orbital with the ionization potential of the system. Using optimally-tuned range-separated hybrid (OT-RSH) functionals, this has been utilized to reliably compute the frontier orbitals' energies for organic molecules in the gas phase,\cite{Stein2010,Refaely-Abramson2011,Refaely-Abramson2012} in bulk crystals \cite{Refaely-Abramson2013,Luftner2014} and at metallic surfaces.\cite{Egger2015,Liu2017} However, no such strict relations are known for deeper lying Kohn-Sham orbitals. 

Owing to these inherent problems with the density functional description, it would be highly desirable to be able to benchmark density functional approaches and also methods that go beyond them, e.g., the $GW$ method within the framework of many-body perturbation theory.\cite{Blase2011,Faber2014,Draxl2014,Marom2017,Golze2019}
Experimentally, the most direct way to measure the valence band electronic structure is ultraviolet photoemission spectroscopy for the occupied states, and inverse photoemission spectroscopy for the unoccupied states.\cite{Kahn2003,Dori2006,Hwang2007,Ueno2008,Heimel2008,Korzdorfer2009} 
Due to photoionization cross section effects, \emph{i.e.}, the dependence of measured spectra on the polarization of the light and the electron emission angle, a one-to-one comparison of such (inverse) photoemission spectra with a computed density of states may, however, not always be conclusive and can sometimes be even  misleading. 
By measuring also the angular distribution of the photoemission intensity, that is, employing angle-resolved photoemission spectroscopy (ARPES), this problem can in fact be turned into an advantage. Over the past decade it has been shown that  
for molecular films adsorbed on (metallic) surfaces, the angular distribution of photoelectrons 
can be simply understood in terms of the Fourier transform of the initial-state orbital. Within this approach, which relies on a plane wave approximation for the final state of the emitted electron and which is now known as photoemission orbital tomography (POT), \cite{Puschnig2009a,Woodruff2016,Puschnig2017}  the photoelectron angular distribution serves as a fingerprint of the orbital structure, \emph{i.e.}, as a momentum-space image of the orbital. This idea has been utilized to deconvolve measured energy distribution curves into contribution from individual molecular orbitals, thereby accessing an experimentally determined density of states \emph{projected} onto molecular orbitals.\cite{Dauth2011,Puschnig2011,Puschnig2016,Zamborlini2017,Yang2019,Kliuiev2019,Brandstetter2020,Saettele2020} 

This projected density of states (pDOS) serves as an extremely rich set of experimental data which is ideally suited to benchmark theoretical electronic structure methods. However, to date the available experimental data has been limited to just a few molecular orbitals, \emph{i.e.}, the frontier $\pi$ orbitals, in a comparably small energy window. In this work, we go considerably beyond these previous investigations, measuring and disentangling the valence band of an adsorbed molecule over a 10~eV energy range below the he Fermi energy ($E_\mathrm{F}$).
Relying on our recent work in which we have shown that POT can also be applied to $\sigma$ orbitals,\cite{Haags2021} we are able to experimentally identify and determine the binding energies and spectral lineshapes of 13 $\pi$ and 12 $\sigma$ orbitals of bisanthene (C$_{28}$H$_{14}$) adsorbed on Cu(110). 
We compare these results with orbital-projected densities of state obtained by DFT employing four widely used exchange-correlation functionals. This allows an orbital-by-orbital assessment of the electronic structure calculations.

\section{\label{sec:comp}Computational Methods}

\subsection{\label{sec:computational}Gas phase calculations}

The geometry and electronic structure of the gas-phase and surface-adsorbed bisanthene was calculated in the framework of DFT. For the former, we have utilized the quantum chemistry package NWChem,\cite{Valiev2010} while for the latter, we have used a repeated slab approach and the VASP program. \cite{Kresse1996,Kresse1996b,Kresse1999}

\begin{table}
\caption{\label{tab:piorbitals} Calculated $\pi$-orbital energies (in eV) of gas phase bisanthene (C$_{28}$H$_{14}$) using different exchange-correlation functionals. The corresponding symmetry labels as well as the number of nodal planes $n$ and $m$ along the $x$ and $y$-directions are listed. }
\begin{ruledtabular}
\begin{tabular}{ccccccc}
symm.  & $n$ & $m$ & PBE  & HSE & PBE0 & B3LYP \\
 \hline
$b_{1u}$ & 4 & 0 & -3.01 & -2.84 & -2.54 & -2.54 \\
$b_{2g}$ & 2 & 3 & -4.00 & -4.28 & -4.62 & -4.39 \\
 $a_{u}$ & 1 & 3 & -5.41 & -5.96 & -6.35 & -6.05 \\
$b_{1u}$ & 2 & 2 & -5.72 & -6.28 & -6.67 & -6.37 \\
 $a_{u}$ & 3 & 1 & -5.87 & -6.43 & -6.82 & -6.52 \\
$b_{3g}$ & 3 & 0 & -6.03 & -6.58 & -6.96 & -6.66 \\
$b_{2g}$ & 0 & 3 & -6.38 & -7.08 & -7.50 & -7.16 \\
$b_{3g}$ & 1 & 2 & -6.99 & -7.75 & -8.17 & -7.82 \\
$b_{2g}$ & 2 & 1 & -7.15 & -7.96 & -8.38 & -8.02 \\
$b_{1u}$ & 2 & 0 & -7.87 & -8.74 & -9.17 & -8.79 \\
$b_{1u}$ & 0 & 2 & -7.97 & -8.88 & -9.31 & -8.92 \\
 $a_{u}$ & 1 & 1 & -8.53 & -9.53 & -9.96 & -9.55 \\
$b_{3g}$ & 1 & 0 & -9.34 & -10.42 & -10.86 & -10.43 \\
$b_{2g}$ & 0 & 1 & -9.44 & -10.55 & -10.99 & -10.55 \\
$b_{1u}$ & 0 & 0 & -10.31 & -11.51 & -11.95 & -11.49 
\end{tabular}
\end{ruledtabular}
\end{table}

\begin{table}
\caption{\label{tab:sigmaorbitals} Calculated $\sigma$-orbital energies (in eV) of gas phase bisanthene (C$_{28}$H$_{14}$) using different exchange-correlation functionals. The corresponding symmetry label as well as the number of nodal planes $n$ and $m$ along the $x$ and $y$-directions are listed. }
\begin{ruledtabular}
\begin{tabular}{ccccccc}
symm.  & $n$ & $m$ & PBE  & HSE & PBE0 & B3LYP \\
 \hline
 $a_{g}$ & 0 & 8 & -7.24 & -8.32 & -8.72 & -8.43 \\
$b_{1g}$ & 7 & 3 & -7.37 & -8.48 & -8.87 & -8.59 \\
$b_{2u}$ & 1 & 8 & -8.10 & -9.25 & -9.66 & -9.34 \\
$b_{2u}$ & 7 & 2 & -8.22 & -9.39 & -9.79 & -9.48 \\
$b_{3u}$ & 4 & 7 & -8.41 & -9.58 & -9.98 & -9.67 \\
$b_{1g}$ & 1 & 7 & -8.66 & -9.86 & -10.26 & -9.94 \\
$b_{3u}$ & 0 & 7 & -8.74 & -9.96 & -10.37 & -10.04 \\
$b_{3u}$ & 2 & 7 & -9.20 & -10.45 & -10.85 & -10.51 \\
 $a_{g}$ & 6 & 2 & -9.28 & -10.53 & -10.94 & -10.60 \\
$b_{1g}$ & 7 & 1 & -9.40 & -10.68 & -11.09 & -10.73 \\
 $a_{g}$ & 4 & 4 & -9.56 & -10.82 & -11.23 & -10.86 \\
 $a_{g}$ & 4 & 6 & -9.95 & -11.25 & -11.65 & -11.28 \\
$b_{2u}$ & 7 & 0 & -10.03 & -11.35 & -11.76 & -11.38 \\
$b_{1g}$ & 3 & 5 & -10.18 & -11.51 & -11.92 & -11.52 \\
$b_{2u}$ & 5 & 6 & -10.22 & -11.53 & -11.95 & -11.56 \\
$b_{1g}$ & 5 & 3 & -10.54 & -11.93 & -12.34 & -11.95 \\
 $a_{g}$ & 0 & 6 & -10.87 & -12.26 & -12.68 & -12.27 \\
$b_{3u}$ & 6 & 1 & -10.93 & -12.32 & -12.74 & -12.33 \\
$b_{2u}$ & 3 & 4 & -11.24 & -12.63 & -13.06 & -12.62 \\
$b_{3u}$ & 2 & 5 & -11.31 & -12.71 & -13.13 & -12.71 \\
 $a_{g}$ & 6 & 0 & -11.48 & -12.91 & -13.33 & -12.90 \\
$b_{2u}$ & 5 & 2 & -11.65 & -13.14 & -13.55 & -13.12 \\
$b_{1g}$ & 1 & 5 & -12.08 & -13.53 & -13.95 & -13.50 \\
$b_{3u}$ & 4 & 3 & -12.59 & -14.18 & -14.60 & -14.12 \\
 $a_{g}$ & 2 & 4 & -12.79 & -14.29 & -14.73 & -14.23 
\end{tabular}
\end{ruledtabular}
\end{table}

\begin{figure}[bt]
\begin{center}
    \includegraphics[width=\columnwidth]{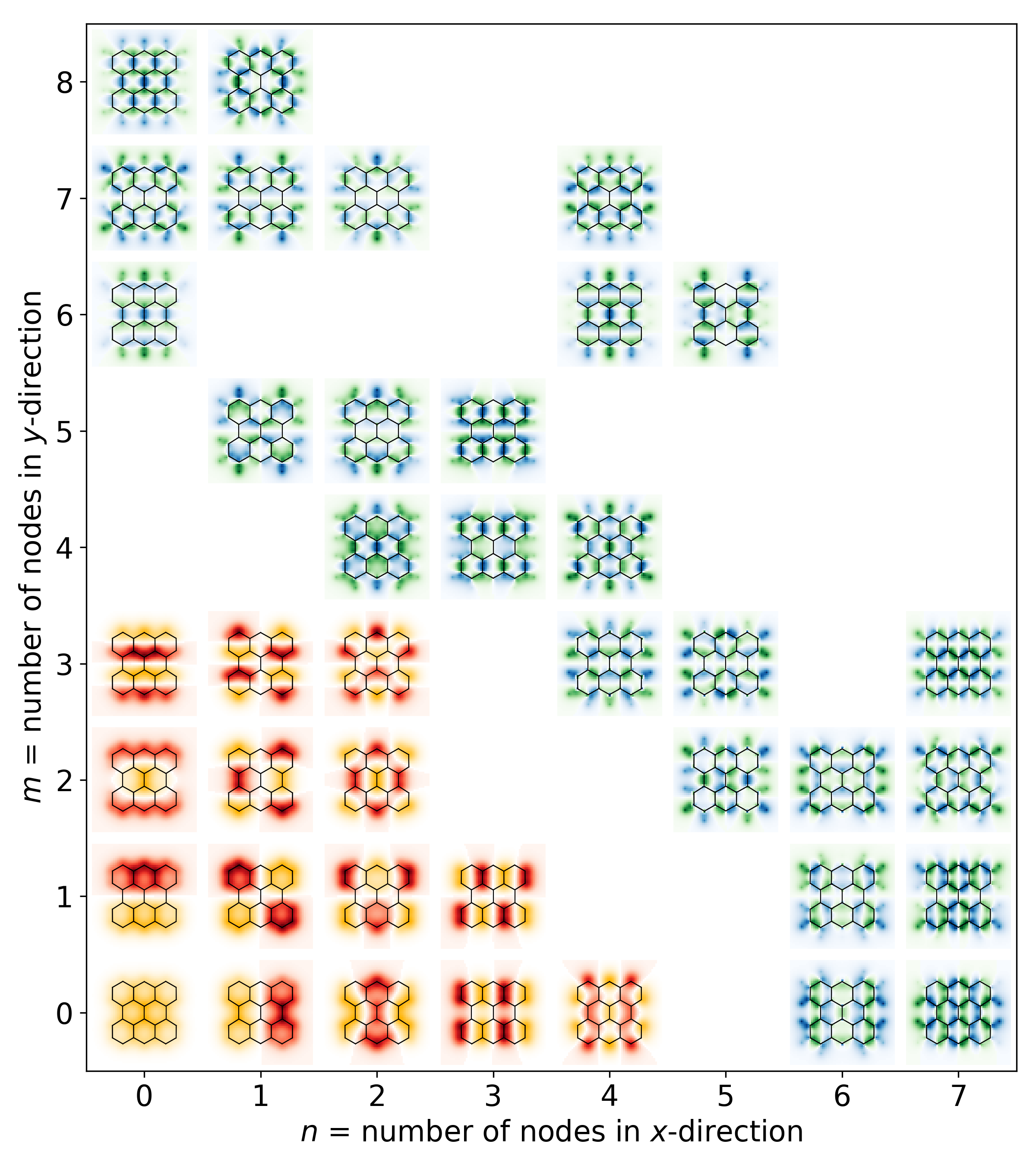}
\end{center}
	\caption{\label{fig:orbitals}  Real space representations of bisanthene's orbitals calculated by using the PBE functional. The red-yellow and blue-green color-scales indicate $\pi$ and $\sigma$ orbitals, respectively. The orbitals are ordered according the number of nodal planes along the $x$ and $y$ directions, defined along the zigzig and armchair edges of the molecule, respectively.}
\end{figure}

All results of the gas-phase calculations including the molecular orbitals and the corresponding momentum maps (see below) are available via a web-based database. \cite{Puschnig2020}
Four different exchange correlation functionals were used: (i) the generalized gradient approximation due to Perdew, Burke and Ernzerhof (PBE),\cite{Perdew1996} (ii) the range-separated hybrid functional due to Heyd, Scuseria and Ernzerhof (HSE),\cite{Heyd2004,Heyd2006} (iii) the global hybrid PBE0 due to Perdew, Ernzerhof, and Burke \cite{Perdew1996a} as well as (iv) the popular B3LYP hybrid functional due to Becke et al. \cite{Becke1993}
The resulting orbital energies of gas-phase bisanthene, which has $D_{2h}$ point group symmetry, are listed in Tables~\ref{tab:piorbitals} and \ref{tab:sigmaorbitals} for $\pi$ and $\sigma$ orbitals, respectively. In addition to the symmetry labels, we have also introduced the two integer numbers $n$ and $m$ which are counting the number of nodes in the respective orbital along the $x$ and $y$ directions, \emph{i.e.}, along the zigzag and armchair edges of bisanthene.
In the remainder of the paper, we will refer to a specific orbital by its type, $\pi$ or $\sigma$, and its number of nodes. For instance,  $\pi_{(4,0)}$ refers to the $\pi$ orbital with $n=4$ and $m=0$ coinciding with the LUMO while, for instance, $\sigma_{(7,3)}$ denotes the $\sigma$ orbital with $n=7$ and $m=3$. 

All 15 $\pi$ and 25 $\sigma$ orbitals listed in Tables~\ref{tab:piorbitals} and \ref{tab:sigmaorbitals}, respectively, are also depicted in Fig.~\ref{fig:orbitals}, classified according to $n$ and $m$. For clarity,  
we use in Fig.~\ref{fig:orbitals}  red/yellow and blue/green to indicate amplitude and phase of the $\pi$ and $\sigma$ orbitals, respectively.
The data presented in Tables ~\ref{tab:piorbitals} and \ref{tab:sigmaorbitals} and in Fig.~\ref{fig:orbitals}  are also available from the online database using the IDs 406 (PBE), 480 (HSE), 482 (PBE0) and 484 (B3LYP).\cite{Puschnig2020}

\subsection{\label{sec:maps}Simulation of momentum maps}

In POT, the momentum-space signatures of molecular orbitals, so-called momentum maps, serve as fingerprints of orbitals which can be used for their identification in ARPES experiments. In this section, we review the underlying theory and assumptions.
In the one-step model of photoemission, the photoemission intensity $I(k_x,k_y;E_\mathrm{kin})$ is given by Fermi's golden rule \cite{Feibelman1974}
\begin{eqnarray}
\label{eq:Feibelman}
I(k_x,k_y;E_\mathrm{kin}) & \propto & \sum_{i}
                 \left| \langle \Psi_f(k_x,k_y;E_\mathrm{kin}) |
                 \ve{A} \cdot \ve{p} | \Psi_i \rangle \right|^2 \nonumber \\
      & \times & \delta \left(E_i + \Phi + E_\mathrm{kin} - \hbar \omega \right).
\end{eqnarray}
Here, $k_x$ and $k_y$ are the components of the emitted electron's wave vector parallel to the surface, which are related to the polar and azimuthal emission angles $\theta$ and~$\phi$,
\begin{eqnarray}
k_x & = & k \sin \theta \cos \phi  \label{eq:kx} \\
k_y & = & k \sin \theta \sin \phi, \label{eq:ky}
\end{eqnarray}
where $k$ is the wave number of the emitted electron, with its kinetic energy being given by $E_\mathrm{kin} = \frac{\hbar^2 k^2}{2m}$, where $\hbar$ is the reduced Planck constant and $m$ is the electron mass. 
The photoemission intensity of Eq.~\ref{eq:Feibelman}  is given by a sum over all transitions from occupied initial states $i$, described by wave functions $\Psi_i$, to the final state $\Psi_f$, characterized by the direction $(\theta,\phi)$ and the kinetic energy of the emitted electron. The $\delta$ function ensures energy conservation, where $\Phi$ denotes the sample work function, $E_i$ the binding energy of the initial state, and $\hbar \omega$ the photon energy. The transition matrix element in Eq.~\ref{eq:Feibelman} is given in the dipole approximation, where $\ve{p}$ and $\ve{A}$, respectively,  denote the momentum operator of the electron and the vector potential of the exciting electromagnetic wave.\cite{Brandstetter2020}

In POT,\cite{Puschnig2017,Brandstetter2020} the final state $\Psi_f$ is commonly approximated by a plane wave (PW). Thereby, the photoemission intensity $I_i$ arising from one particular initial state $i$ turns out to be proportional to the Fourier transform  $\tilde{\Psi}_{i} (\ve{k})$ of the initial state wave function, corrected by the polarization factor $\ve{A} \cdot \ve{k}$,
\begin{equation}
\label{eq:PE1}
I_i(E_\mathrm{kin},k_x,k_y)  \propto \left|\ve{A} \cdot \ve{k}\right|^2  \cdot \left| \tilde{\Psi}_{i} (\ve{k}) \right|^2. 
\end{equation}
A more detailed discussion regarding the applicability of the PW approximation and its limitations can be found in previous publications. \cite{Puschnig2009a,Haags2021}

\begin{figure}[bt]
\begin{center}
	\includegraphics[width=\columnwidth]{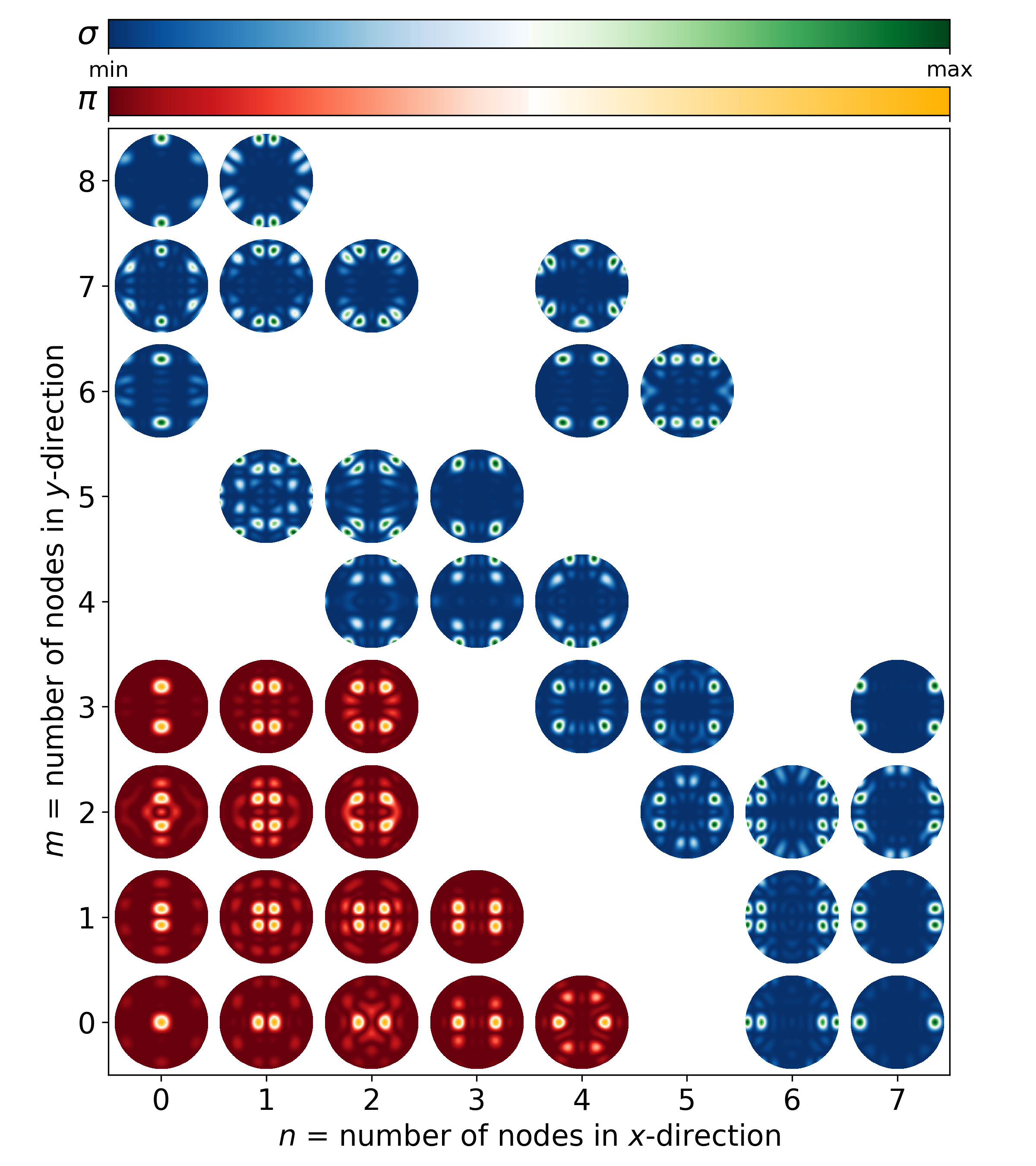}
\end{center}
	\caption{\label{fig:momentummaps}  Momentum space representations of bisanthene's orbitals using the PBE-GGA functional. The red-yellow and blue-green color-scales indicate $\pi$ and $\sigma$ orbitals, respectively. The orbitals are ordered according the number of nodal planes along the $x$ and $y$ directions corresponding to zigzag and armchair edges of bisanthene, respectively. }
\end{figure}

We have computed the momentum maps $I_i(k_x,k_y)$ according to Eq.~\ref{eq:PE1} for all bisanthene orbitals depicted in Fig.~\ref{fig:orbitals} and collected them in Fig.~\ref{fig:momentummaps}, where again the red/yellow color scheme refers to $\pi$ orbitals, while blue/green designates $\sigma$ orbitals. It is evident that the nodal patterns of the orbitals, that is the number of nodes $n$ and $m$ along the two principal directions, is also reflected in the momentum maps which therefore serve as a fingerprints for specific orbitals. This correspondence  will in fact be used in Sec.~\ref{sec:deconvolution} to deconvolve  experimental ARPES data into an orbital-projected density of states.

\subsection{\label{sec:interface}Interface calculations}

For the full bisanthene/Cu(110) interface, we applied the repeated-slab approach. As described previously,\cite{Yang2019} the Cu(110) substrate was modeled with five atomic layers, a lattice parameter of $a = 3.61$~{\AA} and a vacuum layer of at least 17~{\AA} between the slabs to avoid spurious electric fields. \cite{Neugebauer1992} 
The most favorable adsorption site for bisanthene was determined by testing several high-symmetry adsorption sites (hollow, top, short bridge and long bridge) in a local geometry optimization approach, allowing all molecular degrees of freedom and the topmost two Cu-layers to relax until forces were below 0.01 eV/{\AA}. For these geometry optimizations, we have used the PBE exchange-correlation functional\cite{Perdew1996} with the D3 correction for van-der-Waals interactions.\cite{Grimme2010} The projector augmented wave (PAW) method \cite{Bloechl1994,Kresse1999} was employed with a plane wave cutoff of 500 eV and a $3 \times 3 \times 1$ Monkhorst-Pack $k$-point grid with a first-order Methfessel-Paxton smearing of 0.2 eV. 

Based on the relaxed adsorption geometry which turned out to be the short-bridge site, the electronic structure was further analyzed in terms of the molecular orbital-projected density of states (MOPDOS). This MOPDOS was calculated by projecting the Kohn-Sham orbitals of the interacting bisanthene/Cu(110) system onto the orbitals
of the freestanding bisanthene layer, as described in more detail in a previous publication.\cite{Lueftner2017}
Note that for the MOPDOS analysis, we employed the same set of exchange-correlation functionals already used for the gas-phase calculations, that is, (i) PBE \cite{Perdew1996}, (ii) HSE \cite{Heyd2004,Heyd2006}, (iii) PBE0\cite{Perdew1996a} and (iv) B3LYP.\cite{Becke1993}

\section{\label{sec:exp}Experimental Methods}

\subsection{Sample preparation}

Our experiments were performed in ultra-high vacuum ($\approx 10^{-10}$ mbar). The Cu(110) single crystal
was cleaned by several cycles of sputtering by Ar$^+$ ions at 1 keV and subsequent annealing at 800 K. 
A film of the 10,10'-dibromo-9,9'-bianthracene  precursor (Sigma-Aldrich, CAS
number 121848-75-7) was deposited by evaporation from a molecular evaporator (Kentax GmbH)
onto the crystal surface held at room temperature. Subsequently, the sample was annealed at 525 K to trigger the chemical reaction as described elsewhere.\cite{Yang2019}

\subsection{Photoemission experiments}

\begin{figure*}[bt]
\begin{center}
l	\includegraphics[width=0.75\textwidth]{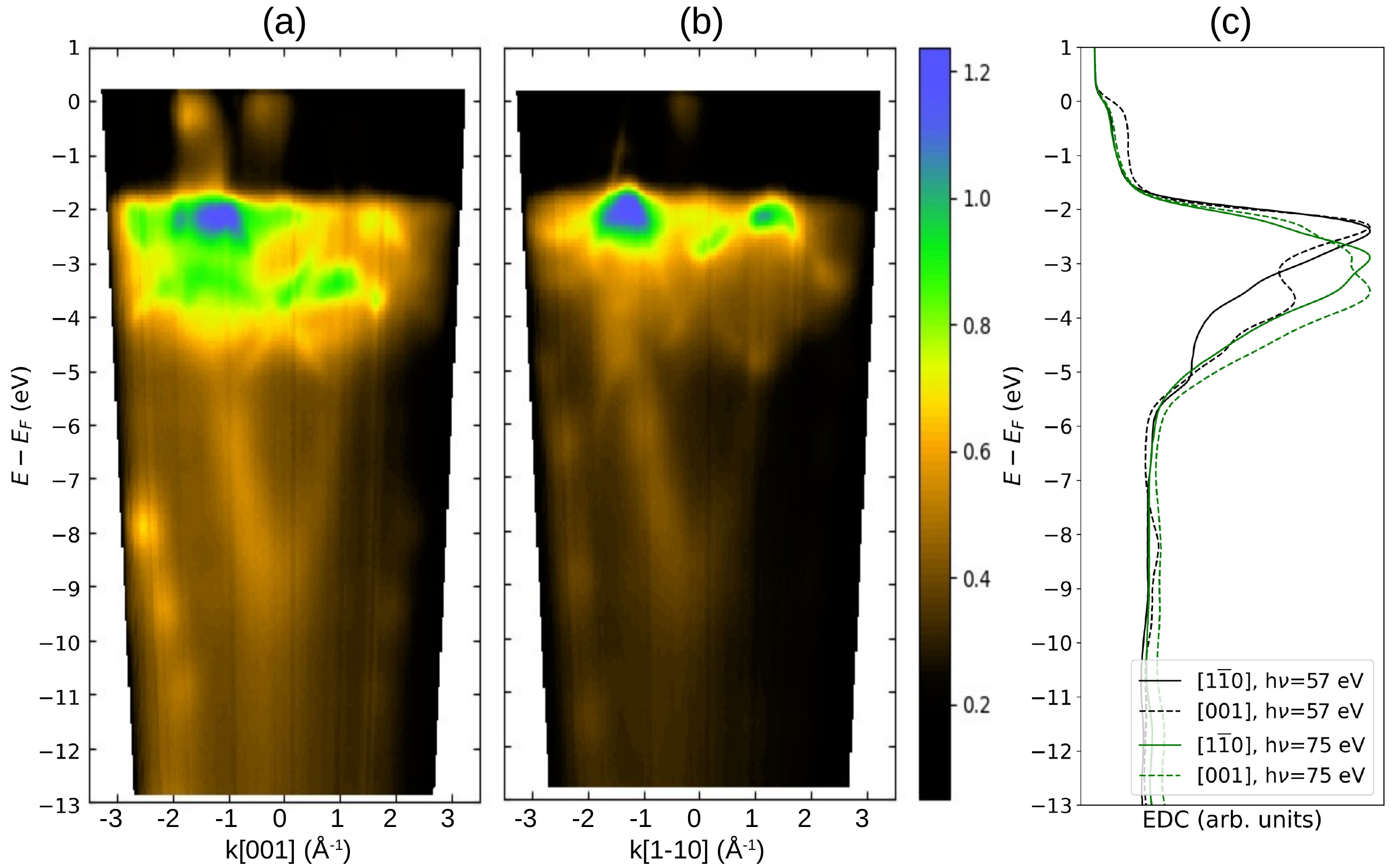}
\end{center}
	\caption{\label{fig:expedc}  (a, b) Experimental  $k_\parallel$-resolved band maps for bisanthene/Cu(110) measured with a photon energy of 45 eV along the  $[001]$ and $[1\overline{1}0]$ azimuths, respectively. (c) Experimental $k_\parallel$-integrated energy distribution curves (EDCs) for bisanthene/Cu(110) obtained with photon energies of 57 eV (black lines) and 75 eV (green lines). The full (dashed) lines depict the photoemission intensities integrated over all $k_\parallel$ emissions along the $[1\overline{1}0]$ ($[001]$) direction. }
\end{figure*}

Photoemission experiments were conducted at the Metrology Light Source insertion device
beamline of the Physikalisch-Technische Bundesanstalt (Berlin, Germany).\cite{Gottwald2019} 
$p$-polarized ultraviolet light with an incidence angle of 40$^\circ$ to the surface normal was used. 
In this geometry, the $\ve{A} \parallel \ve{k}$ condition, where $\ve{A}$ is the vector potential of the incident light and $\ve{k}$ the wave vector of the photoelectrons, is approximately fulfilled for most molecular emissions in forward direction which is favorable for applying the plane wave approximation for the final state.\cite{Puschnig2017}

Two different types of photoemission experiments were conducted using a toroidal electron analyzer.\cite{Broekman2005}
First, we measured experimental band maps, \emph{i.e.}, the photoemission intensity  $I$ over a larger binding energy window of 13 eV for two emission planes along the principal azimuths of the Cu(110) substrate, by recording emission angles ranging from -85$^\circ$ to +85$^\circ$. After conversion to parallel momentum components $k_x$ and $k_y$, respectively, these band maps,  $I_\mathrm{exp}(E_b, k_x)$ and $I_\mathrm{exp}(E_b, k_y)$, along the $[001]$ ($x$) and $[1\overline{1}0]$ ($y$) directions of Cu(110) are depicted in panels (a) and (b) of Fig.~\ref{fig:expedc}. They prove that we are indeed able to observe emissions at high binding energies that originate from the molecule, and already gives an indication about the orbital energies of deeper lying molecular states. 
Note that these band maps were obtained with a photon energy of 45 eV, while the angle-integrated energy distribution curves (EDCs) shown in panel (c) were recorded with 57 eV (black lines) and 75 eV (green lines). For energies between $-2$ and $-5$~eV below the Fermi energy, the spectra in Fig.~\ref{fig:expedc}c are dominated by substrate emissions from Cu $d$ states. 
Outside this energy window, only little structure is visible in the EDCs, despite the anisotropy due to the azimuth of the emission plane, $[001]$ or $[1 \overline{1} 0]$, respectively. This underlines the need for angular resolution in the band maps, where the molecular emission patterns become much more evident. 
In the band maps of Fig.~\ref{fig:expedc}a and b, one can indeed clearly assign emissions above the Cu $d$  band that are  visible only in the $[001]$ direction, and two distinct bands of emissions below the Cu $d$ band originating from the $\pi$ and $\sigma$ orbitals of bisanthene. It is important to note that the emission signatures of the $\sigma$ orbitals appear at large $k_\parallel$ values, therefore high enough photon energies are needed to allow for a sufficiently large photoemission horizon. On the other hand, since at higher photon energies, the cross section for photoionization drops and also the mean free path of photoelectrons from the bulk, including those which are inelastically scattered,  increases, a good compromise is found around $\approx 50$~eV photon energy to resolve the $\sigma$ orbitals.

Despite the rich information in the band maps, it is clear that, because of the nodal structure of the orbitals which is also reflected in the momentum maps (see Fig.~\ref{fig:momentummaps}), many molecular orbitals will not be visible along the principal azimuths of the substrate. Thus, in the second type of ARPES experiment, we measured momentum maps. Here, at fixed binding energies $E_b$, full $\ve{k}_\parallel$ intensity maps covering the entire half-space above the sample were obtained. This leads to a three-dimensional data cube $I_\mathrm{exp}(E_b, k_x, k_y)$, \emph{i.e.}, the intensity of photoemission as a function of binding energy $E_b$ and the parallel momenta components $k_x$ and $k_y$, respectively. Using the toroidal electron energy analyzer, these momentum maps were recorded by collecting the electrons in a given emission plane by rotating the sample around its normal in 1$^\circ$ steps. In this way, the full photoemission intensity distribution in the $\ve{k}_\parallel$-plane perpendicular to the sample normal was determined. Such momentum maps for bisanthene/Cu(110) have been analyzed previously in the low binding energy range, where they have revealed fingerprints of $\pi_{(4,0)}$ (filled LUMO), $\pi_{(2,3)}$ (HOMO) and $\pi_{(1,3)}$ (HOMO-1), \cite{Yang2019} and for selected binding energies below the Cu $d$ band, where they have shown the emissions around $-5.2$~eV to originate from the $\sigma_{(7,3)}$, the $\sigma_{(0,8)}$ and the $\pi_{(0,3)}$ orbitals, respectively. \cite{Haags2021}

\subsection{\label{sec:deconvolution}Deconvolution of experimental ARPES data}

In the orbital deconvolution procedure,\cite{Puschnig2011,Brandstetter2020} we make use of the energy and momentum dependence of the data cube $I_\mathrm{exp}(E_b, k_x, k_y)$ to deconvolve experimental data into individual orbital contributions. This provides an orbital-by-orbital decomposition of the
experimental data cube into orbital-projected densities of states (pDOS), that can be readily compared to the computed MOPDOS.
Specifically, the deconvolution of the experimental data cube consists of minimizing the squared differences between the experimental and simulated momentum maps, 
\begin{widetext}
\begin{equation}
\label{eq:chi2a}
\chi^2(w_1,w_2,\cdots, w_N) = \sum_{k_x,k_y} \left[ I_\mathrm{exp}(k_x,k_y,E_b) - \sum_{i=1}^N w_i(E_b) I_i(k_x,k_y) \right]^2 
\end{equation}
\end{widetext}
by adjusting the $N$ weights $w_i$ of all orbitals $i$ with the simulated momentum maps $I_i(k_x,k_y)$ that are allowed to contribute to the measurement data. Since the minimization is performed for each binding energy $E_b$ separately, one thereby obtains 
a set of orbital-projected densities of states given by the weight functions $w_i(E_b)$. 

\begin{figure}[bt]
\begin{center}
	\includegraphics[width=\columnwidth]{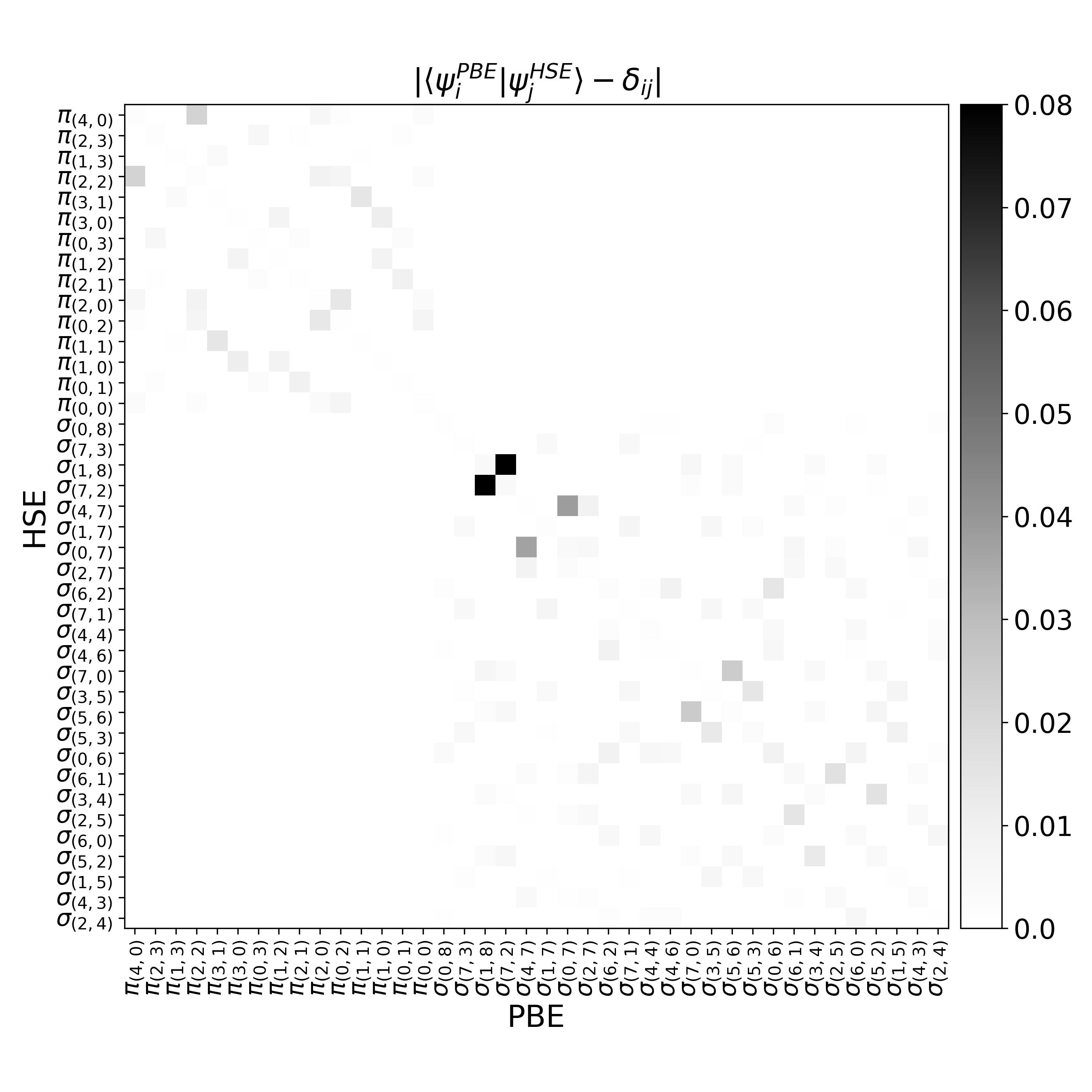}
\end{center}
	\caption{\label{fig:overlaps}  Deviation of the overlap matrix between bisanthene's $\pi$ and $\sigma$ orbitals computed with PBE and HSE functionals, respectively, from the identity matrix according to Eq.~\ref{eq:overlap}.}
\end{figure}

From Eq.~\ref{eq:chi2a}, it is clear that the so-obtained experimental pDOS will also depend on the set of simulated momentum maps $I_i(k_x,k_y)$ that are used in the deconvolution procedure. It is therefore important to check how sensitive these computed momentum maps are with respect to the choice of the exchange-correlation (xc) functional. We found that the momentum maps are robust and remain almost unaffected by the choice of the xc functional. As an example, this is illustrated for the PBE and HSE functionals in Fig.~\ref{fig:overlaps}, which illustrates how much the overlap matrix of PBE and HSE orbitals deviates from the identity matrix, thus we are plotting the quantity
\begin{equation}
\label{eq:overlap}
\left| \left\langle \Psi_i^\mathrm{PBE} \right| \left. \Psi_j^\mathrm{HSE} \right\rangle  - \delta_{ij} \right|.
\end{equation}
Using a grey-scale density map, we note the maximum deviation from perfect overlap to be less than 9\%,  in most cases, however, the similarity between the PBE and HSE orbitals is much better and the resulting momentum maps from these two functionals are essentially  indistinguishable. Also note that a similar agreement is found when comparing the PBE orbitals with those of the other two xc functionals studied in this work.

\section{\label{sec:results}Results and Discussion}

\subsection{\label{subsec:dos}Density of states}
 
\begin{figure}[bt]
\begin{center}
	\includegraphics[width=\columnwidth]{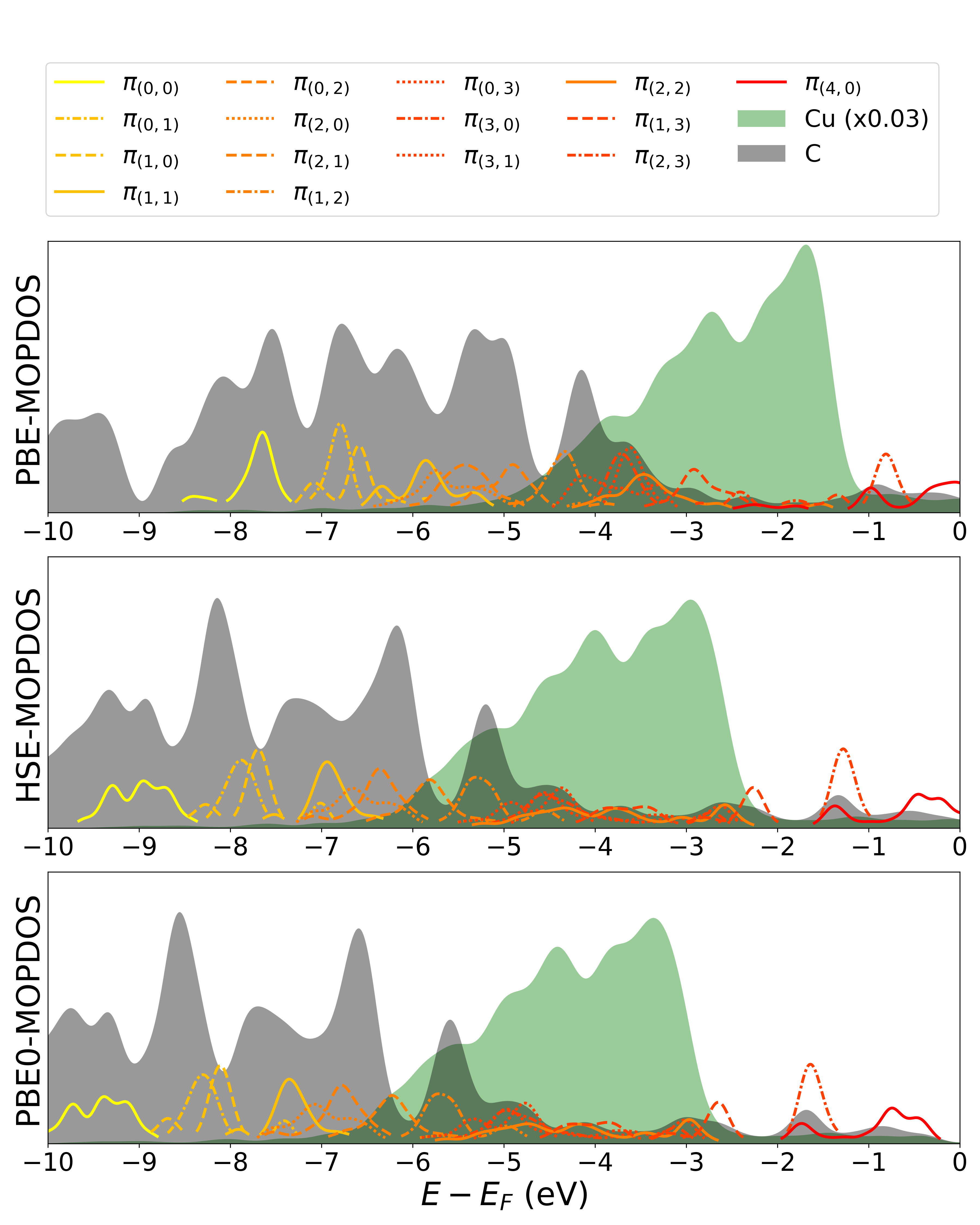}
\end{center}
	\caption{\label{fig:mopdos_pi}  Computed projected density of states for bisanthene/Cu(110) for three different exchange-correlation functionals: PBE-GGA (top), HSE (middle) and PBE0 (bottom). Red-yellow color-coded curves correspond to the MOPDOS for all $\pi$ orbitals of bisanthene. Green and grey shaded areas depict the DOS projected on copper and carbon atoms, respectively.}
\end{figure}

We start the analysis of the electronic structure of the bisanthene/Cu(110) interface by comparing the density of states as obtained from various exchange-correlation functionals. Fig.~\ref{fig:mopdos_pi} depicts the computational results for PBE (top panel), HSE (middle panel) and PBE0 (bottom panel). 
The comparison reveals several interesting trends. First, we notice that the energetic position of the copper $d$ band (green shaded areas) depends sensitively on the xc-functional. For PBE, its onset is around $1.2$~eV below $E_\mathrm{F}$, while the hybrid functionals HSE and PBE0 yield significantly deeper $d$ band positions of about $-2.3$ and $-2.7$~eV, respectively. Thus, the inclusion of exact exchange leads to an improvement when comparing to the experimentally observed onset at about $-2.0$~eV (compare Fig.~\ref{fig:expedc}). However, this correction overshoots somewhat for the range-separated HSE and even more so for the global hybrid PBE0. It should also be noted that the computed $d$ bandwidth of about 3.7 eV remains almost unchanged for all exchange-correlation functionals tested in this work and that this value is almost 25\% larger than the experimentally observed bandwidth of 3~eV. 

Similar trends are also observed when comparing the densities of states (DOS) originating from the adsorbed organic molecule. To this end, we have projected the DOS of the bisanthene/Cu(110) system onto the carbon atoms of bisanthene (grey shaded areas) and also computed the MOPDOS for all 15 $\pi$ orbitals introduced in Table~\ref{tab:piorbitals} (red/yellow colored lines). 
Note that the overall bandwidth of occupied $\pi$ bands shows some variation with type of xc functional. For PBE, the deepest $\pi$ orbital, $\pi_{(0,0)}$, peaks around $-7.7$ eV, while the onset of the $\pi$ bands for the hybrid functionals is at $-9.3$~eV (HSE) and $-9.8$~eV (PBE0), which should be compared with the experimental value of $-8.7$~eV as determined from the band map data shown in Fig.~\ref{fig:expedc} evaluated at normal emission. 
Thus, among the three tested functionals, HSE again performs best. 
The choice of the functional also affects the calculated amount of charge transfer into the LUMO ($\pi_{(4,0)}$) and influences the degree of hybridization between molecular and metallic states. 
For PBE, the LUMO is partially occupied, for HSE it is almost entirely below $E_\mathrm{F}$, while for PBE0 it is fully occupied. 
A pronounced difference of PBE compared to HSE and PBE0 can, for instance, also be observed for the HOMO-1 ($\pi_{(1,3)}$). 
Due to the high-lying $d$ band in PBE, $\pi_{(1,3)}$ overlaps entirely with the $d$ band and PBE predicts a strong hybridization, as indicated by the broad MOPDOS curve.
The hybridization is predicted to be less strong in HSE and PBE0, which both yield a $\pi_{(1,3)}$-related MOPDOS peak right above the copper $d$ band, in agreement with the experimental observation. \cite{Yang2019}

\begin{figure}[bt]
\begin{center}
	\includegraphics[width=\columnwidth]{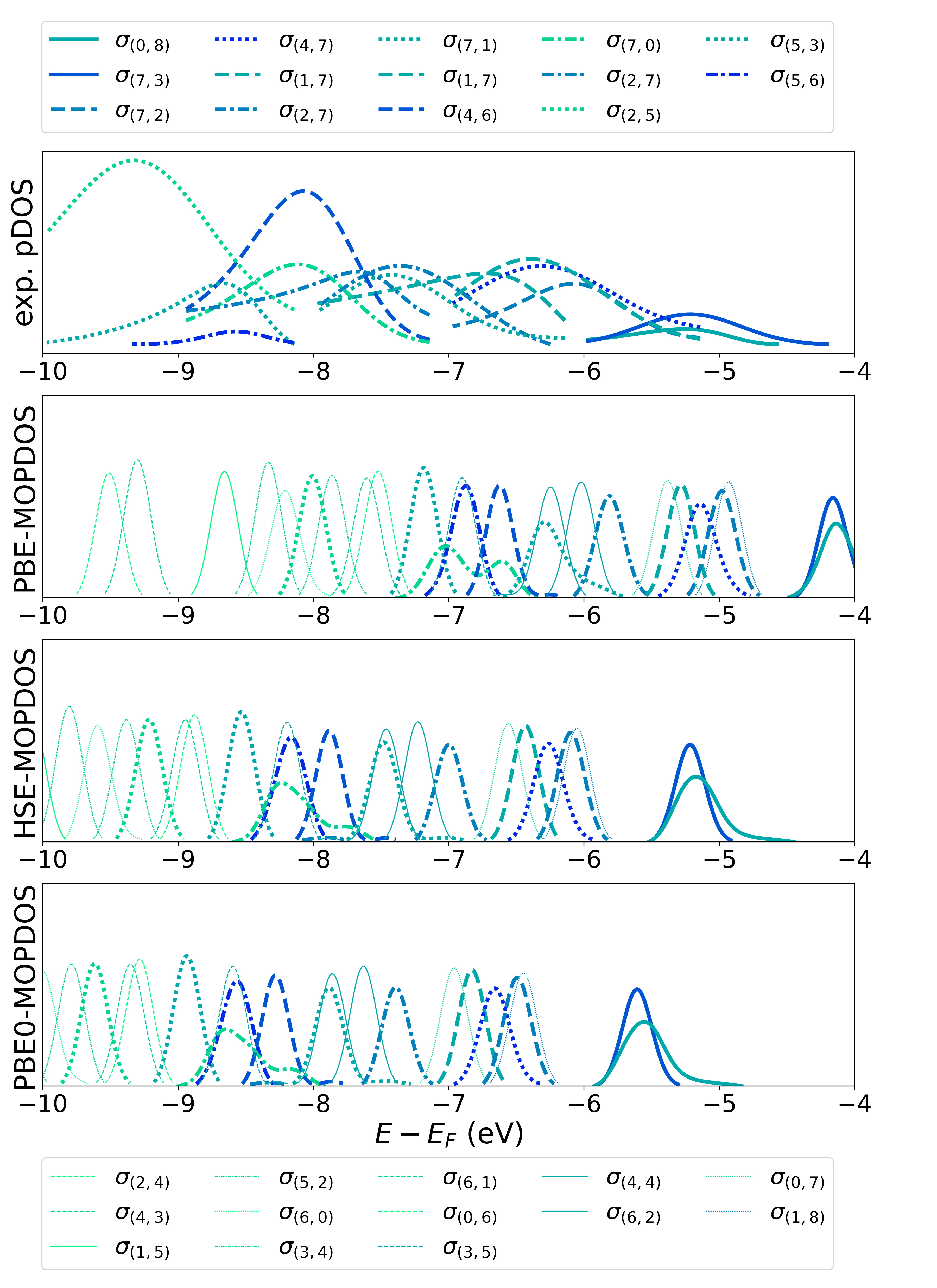}
\end{center}
	\caption{\label{fig:deconvolution_vs_MOPDOS}  Comparison of the experimental pDOS for 14 $\sigma$ orbitals according to Eq.~\ref{eq:chi2a} (top panel) with the computed MOPDOS for the three exchange-correlation functionals PBE, HSE and PBE0.}
\end{figure}

We further deepen our understanding of the electronic structure of the bisanthene/Cu(110) interface by comparing in Fig.~\ref{fig:deconvolution_vs_MOPDOS} the MOPDOS for $\sigma$ orbitals with respective experimental pDOS data as obtained from the deconvolution of the momentum maps according to Eq.~\ref{eq:chi2a}. 
In Ref.~\onlinecite{Haags2021}, we demonstrated that the energetic positions and the appearance of the pDOS curves of the two topmost $\sigma$ orbitals, namely $\sigma_{(0,8)}$ and $\sigma_{(7,3)}$, can be used to distinguish bisanthene from a possible metalated molecular species. 
Here, we extend the analysis of the experimental data considerably and deconvolve the experimental data cube $I_\mathrm{exp}(E_b, k_x, k_y)$ into individual contributions from 12 different $\sigma$ orbitals of bisanthene, spanning an energy range from about $5$ to $10$ eV below the Fermi energy (see top panel of Fig.~\ref{fig:deconvolution_vs_MOPDOS}). 
The lower three panels of this figure display the corresponding calculated MOPDOS of the $\sigma$ orbitals as obtained from PBE, HSE and PBE0, respectively. 
It should be noted that in addition to the 12 $\sigma$ orbitals for which experimental pDOS curves have been obtained (thick lines), further 13 $\sigma$ orbitals are included in the theoretical MOPDOS plots (thin lines) which could not be identified in the experimental data cube.

Regarding the influence of the exchange-correlation functional, we recognize the same trend as for the $\pi$ orbitals. 
There is an overall shift to larger binding energies when going from PBE to HSE and to PBE0, accompanied by an increase in the bandwidth. 
Also in this respect, the HSE functional shows the best agreement with the experimental pDOS: the peak positions of the two topmost $\sigma$ orbitals, $\sigma_{(0,8)}$ and $\sigma_{(7,3)}$, at about $-5.2$~eV are in almost perfect agreement with the experimental energy location, and also the lowest $\sigma$ orbital that could be observed experimentally, $\sigma_{(5,3)}$ at $-9.3$~eV, appears very close to the HSE peak position of $9.2$~eV. 
On the other hand, all PBE $\sigma$ energy positions are too small, as is the total width of the $\sigma$ band, while PBE0 slightly overestimates the binding energies and total width of the $\sigma$ band. 

In addition to the peak positions, it is also important to inspect the widths of the individual peaks, both in the experimental pDOS as well as in the calculated MOPDOS. 
Starting with the theoretical MOPDOS, we notice little variation over all $\sigma$ orbitals and almost no influence of the functional, with an overall full width at half maximum (FWHM) of approximately 0.25 eV. 
This is to be contrasted with the considerably larger FWHM in the theoretical MOPDOS of the $\pi$ orbitals (Fig.~\ref{fig:mopdos_pi}). 
Due to the  weaker spatial overlap of the $\sigma$ orbitals with the substrate states, the hybridization is weaker as compared to the $\pi$ orbitals, which  overlap more strongly with the substrate and hence exhibit an enhanced tendency to hybridize with the latter.
When inspecting the lineshape of individual orbitals in the experimental pDOS curves, however, we observe a FWHM in the range from $1.0$ to even $1.3$ eV for the $\sigma$ orbitals, much larger than the corresponding widths in the theoretical MOPDOS, but also significantly larger than the experimental resolution of the toroidal electron energy analyzer. 
To explain this discrepancy, we remark that the experimental pDOS is inextricably interweaved with the underlying photoemission process, while theoretical MOPDOS reflects a pure density of states. 
Specifically, we suggest that the reason for the much larger FWHM in the experimental data arises from the short lifetime of the photohole, which leads to a broadening of the spectral signatures. 
Theoretically, such an effect would be contained in Eq.~\ref{eq:Feibelman} if one were to replace the $\delta$ function by the spectral function, for instance, from a $GW$ calculation.\cite{Gerlach2001,Marini2002,Yi2010} While a calculation of the spectral function is beyond the scope of the present work, we note that the experimentally observed peak width would suggest extremely short lifetimes of only 0.6 fs. This estimate may appear surprising but we note that the energy range of the $\sigma$ orbitals, $-5$ to $-10$~eV, is considerably below the copper $d$ band and that quasiparticle calculations for Cu have predicted hole lifetimes of only $\approx 1$~fs for states 5 eV below $E_\mathrm{F}$.\cite{Yi2010}

\subsection{Level alignment}

\begin{figure*}[bt]
\begin{center}
	\includegraphics[width=\textwidth]{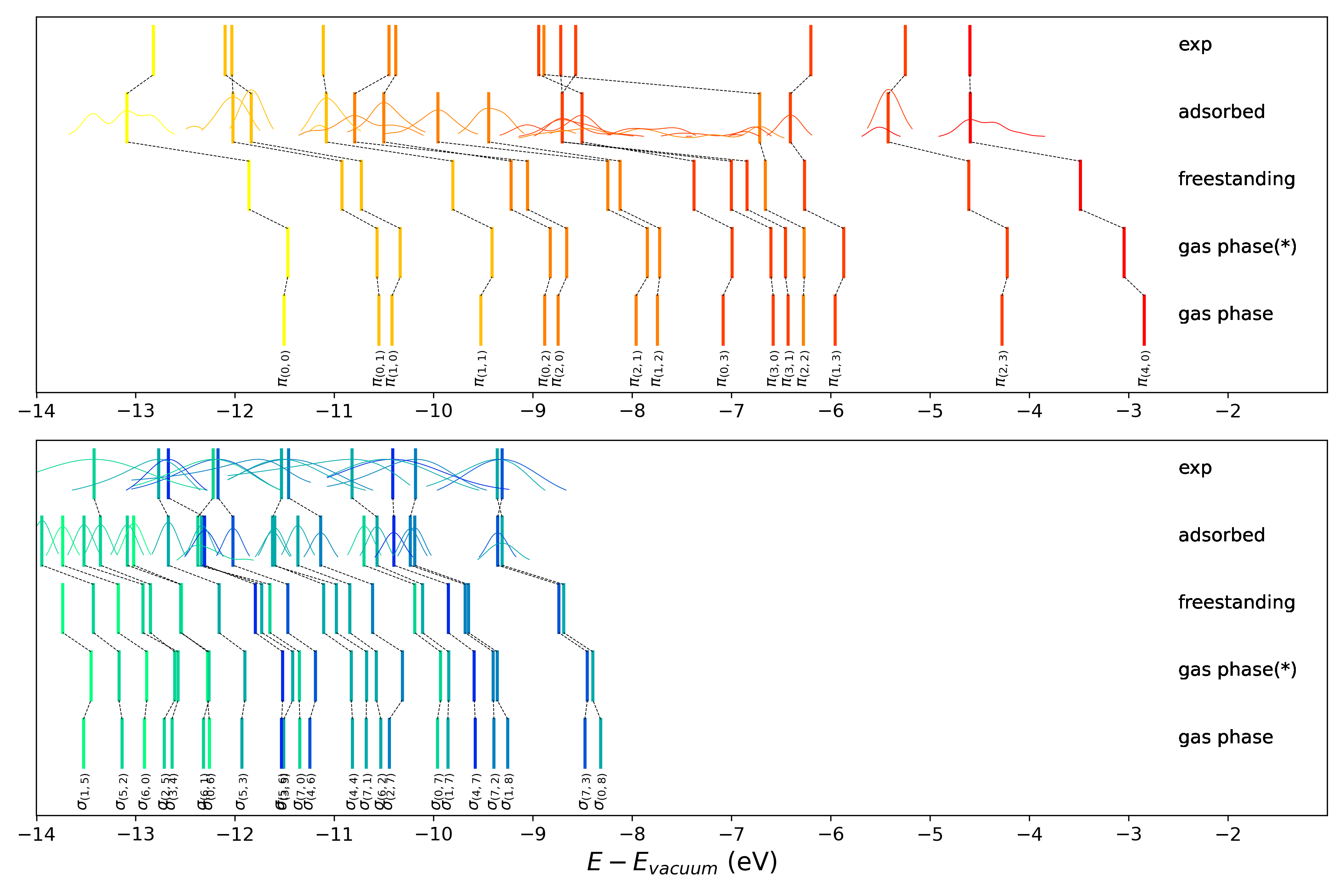}
\end{center}
	\caption{\label{fig:HSE_energies}  Experimental (row: exp) and DFT orbital energies using the HSE functional (rows: adsorbed, freestanding, gas-phase(*) and gas-phase) for $\pi$ orbitals (top panel) and $\sigma$ orbitals (bottom panel) of bisanthene. See text for more details.}
\end{figure*}
	
The goal of this section is to disentangle the effects that lead to the energy level alignment of the molecular states with the states of the metal  upon adsorption on the metal surface. 
To this end, we dissect the adsorption process into several steps. 
First, we start with the bisanthene molecule in the gas phase in its native $D_{2h}$ point group symmetry. 
The respective orbital energies are presented in Tables~\ref{tab:piorbitals} and \ref{tab:sigmaorbitals} for the $\pi$ and $\sigma$ orbitals, and are reproduced in the row labeled ''gas phase'' in Fig.~\ref{fig:HSE_energies} for the HSE functional. 
Note that in this figure all orbital energies are referenced to the vacuum level. 
Second, we take into account the structural distortion of the molecule as predicted by our van-der-Waals-corrected GGA optimizations of adsorbed bisanthene, \cite{Yang2019} but compute orbital energies the molecule in the gas phase. 
The resulting HSE orbital energies are shown in Fig.~\ref{fig:HSE_energies} in the row labeled ''gas-phase(*)''. 
Overall, the geometry-induced level shifts are small (in the order of $0.1$~eV) with the exception of $\pi_{(4,0)}$, which is the partially occupied former LUMO, and therefore experiences also a somewhat larger energy stabilization of about 0.2~eV due to the adsorption-induced geometrical changes. 

In the third step, we calculate a freestanding layer of bisanthene molecules, \emph{i.e.}, we take the relaxed geometry of the bisanthene/Cu(110) interface and  cut away the Cu(110) substrate, but keep the distorted structure of the molecules frozen. 
The resulting orbital energies are labeled ''freestanding'' in Fig.~\ref{fig:HSE_energies}. 
It is important to note that we apply periodic boundary conditions in this ''freestanding'' calculation and, consequently, each molecular state develops into a band. 
However, the intermolecular interactions are comparably small and the concomitant bandwidths are only in the order of 0.1~eV. 
The vertical lines plotted in Fig.~\ref{fig:HSE_energies} therefore represent the center of these bands. 
The main result of forming a freestanding layer of molecules of ''gas-phase(*)'' molecules is an overall shift of about $0.25-0.35$~eV to lower energies, essentially independent of the specific orbital. 
This shift originates from a step in the vacuum potential of the freestanding layer, that arises from a geometry-induced dipole perpendicular to the molecular plane, sometimes also referred to as ''bending dipole''.\cite{Willenbockel2012} 
In fact, the infinitely extended freestanding layer of bisanthene has two vacuum potentials depending on whether one removes the electron in positive or negative $z$ direction. 
This step in the vacuum potential amounts to about 0.25~eV and is caused by the concave shape of the adsorbed bisanthene, with the terminating hydrogens at the zig-zag edges of the molecule displaced by about 0.15~{\AA} upwards compared to the central carbon atoms.
		
Finally, we investigate how the orbital energies are further affected when the freestanding layer of molecules is brought into contact with the Cu(110) surface. 
These results of this last step of our gedanken experiment are depicted in the row labeled ''adsorbed'' and should  also be directly compared with the experimental observations (row ''exp''). 
The thin lines in the row ''adsorbed'' are in fact the MOPDOS curves already presented in Figs.~\ref{fig:mopdos_pi} and \ref{fig:deconvolution_vs_MOPDOS} for the $\pi$ and $\sigma$ orbitals, respectively, while, the thick vertical lines are drawn at the global maxima of the respective MOPDOS curve. 
When inspecting the changes from ''freestanding'' to ''adsorbed'', first for the $\sigma$ orbitals, we notice an overall shift of roughly 0.5~eV to lower energies. 
This shift is due to the so-called bonding dipole, a dipole and the concomitant step in the potential due to adsorption-induced charge density rearrangements.\cite{Willenbockel2012}  
As mentioned before, the final energy positions obtained from the HSE for the full bisanthene/Cu(110) interface are in good agreement with the experimental values as obtained from the orbital deconvolution procedure.

The same effect, namely an overall energy shift of 0.5~eV to lower energies, also affects the $\pi$ orbitals. 
However, the $\pi$ orbitals are additionally stabilized due to the hybridization with the underlying Cu atoms.
This latter effect can, for instance, be observed for $\pi_{(4,0)}$ and $\pi_{(2,3)}$, \emph{i.e.}, the LUMO and HOMO, respectively, which exhibit shifts of 1.1 and 0.8~eV, respectively. 
Subtracting the shift of 0.5~eV from the overall bonding dipole, we are left with orbital-specific bond stabilizations of roughly 0.6 and 0.3~eV for the LUMO and HOMO, respectively. 
It must be noted, however, that this analysis is somewhat subjective because of the ambiguity of assigning a \emph{single} energy position to each molecular orbital of the adsorbed system, while, more precisely, molecular resonances of finite width are formed. 
This is particularly true for those $\pi$ states which overlap energetically with the Cu $d$ band between $-6$ and $-10$~eV below the vacuum level. 
For instance, the HOMO-1 and HOMO-2 ($\pi_{(1,3)}$ and $\pi_{(2,2)}$), although having their MOPDOS maxima at the low binding energy side of the $d$ band, are spread out over the whole $d$ band region, making a definite assignment to a single binding energy questionable.
Below the copper $d$ band, however, for instance in the case of the lowest lying $\pi_{(0,0)}$, we can again understand the energy shift of 1.25~eV from ''freestanding'' to ''adsorbed'' as an overall shift of 0.5~eV due to the bonding dipole potential step and an additional bond stabilization of 0.75~eV owing to the molecule-substrate bond.
	
\subsection{Quantitative comparison of functionals}

\begin{figure}[bt]
\begin{center}
	\includegraphics[width=\columnwidth]{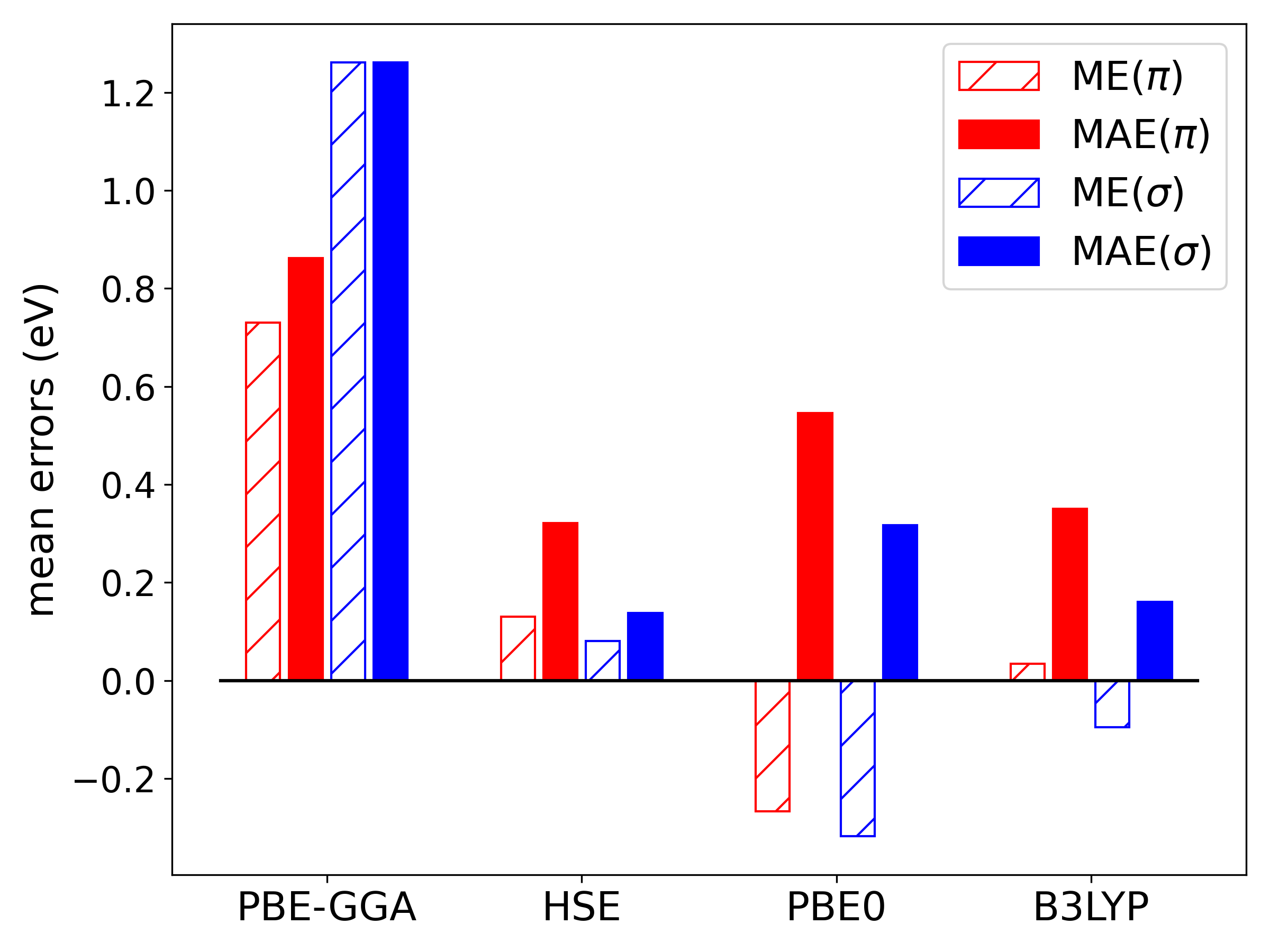}
\end{center}
	\caption{\label{fig:errors} Mean errors (ME) and mean absolute errors (MAE) for $\pi$ (red) and $\sigma$ (blue) orbital energies as obtained using four different exchange-correlation functionals.}
\end{figure}	
	
We now return to the issue of quantifying the performance of commonly applied exchange-correlation functionals for predicting the energy level alignment at organic/metal interfaces, making use of the comprehensive experimental information gained by photoemission orbital tomography. 
Some general trends regarding orbital energies computed with the help of the three functionals PBE, HSE and PBE0 have already been discussed in Sec.~\ref{subsec:dos}.
In the following, we quantify their accuracy by computing mean errors (ME) and mean absolute errors (MAE) defined as follows
\begin{eqnarray}
\mathrm{ME} & = & \frac{1}{N} \sum_i (E_i^\mathrm{exp} - E_i^\mathrm{DFT}) \label{eq:ME}\\
\mathrm{MAE} & = & \frac{1}{N} \sum_i | E_i^\mathrm{exp} - E_i^\mathrm{DFT} | \label{eq:MAE}.
\end{eqnarray} 
Here, $E_i^\mathrm{exp}$ is the binding energy of orbital $i$, measured with respect to the vacuum level, as determined from the orbital deconvolution using the photoemission data cube. 
These values are shown in Fig.~\ref{fig:HSE_energies} in the rows labeled ''exp'' for the $\pi$ and $\sigma$ orbitals, respectively. 
For the case of the HSE functional, the theoretical values, $E_i^\mathrm{DFT}$, are also marked in Fig.~\ref{fig:HSE_energies} as thick vertical lines in the rows denoted ''adsorbed''. 
Note that we have evaluated Eqs.~\ref{eq:ME} and \ref{eq:MAE} separately for the $\pi$ and $\sigma$ orbitals.
Accordingly, $N$ denotes the number of orbitals included in the summation.
The results are displayed in the bar diagram of Fig.~\ref{fig:errors}, where red and blue colors represent the errors for $\pi$ and $\sigma$ orbitals, respectively, and hatched and filled bars denote mean errors and mean absolute errors.

The largest errors are found for the generalized gradient approximation PBE, for which we observe also a significant difference between the $\pi$ and $\sigma$ orbitals. 
The reason that PBE errors for the $\sigma$ orbitals are about 0.4~eV larger than for $\pi$ orbitals can be attributed to the self-interaction error: the $\sigma$ orbitals are more localized than $\pi$ orbitals and therefore more vulnerable to self-interaction errors. 
This self-interaction error is clearly mitigated in the hybrid functionals by incorporating a fraction $\alpha$ of Hartree-Fock (HF) exchange.
Note that $\alpha$ is 0.25 for HSE and PBE0 and 0.2 for B3LYP. 
Among these three hybrid functionals, HSE performs best, followed by B3LYP and PBE0. 
The fact that in the range-separated HSE functional HF exchange is only included in the short range, apparently outperforms the global hybrid PBE0. 
The cause might be that, in the language of optimally-tuned range separated hybrid functionals,\cite{Refaely-Abramson2013,Luftner2014,Kronik2018} the HSE functional can be viewed as having effectively infinite dielectric screening in the long-range, which seems appropriate for the bisanthene/Cu(110) system studied in this work. 
For other molecule/metal interfaces with larger molecule-metal distances, for instance on Ag or Au surfaces, the superior performance of HSE over PBE0 may therefore not hold in general. 
Finally, the results of the global hybrid functional B3LYP are only slightly worse than those of HSE. 
Presumably, its smaller fraction of HF exchange partly corrects for the overshooting in the ME of PBE0 and compensates for the long-range screening of HSE. 
Finally, we stress that these findings are system-dependent and expected to vary with the type of substrate and the molecule-substrate distance.

\section{\label{sec:conclusion}Conclusion}

Using the example of bisanthene/Cu(110), we have shown that photoemission orbital tomography (POT) is able to provide experimental data on the orbital binding energies of individual molecule-derived states at the interfaces between organic molecules and metallic surfaces. 
The so-obtained binding energies of 13 $\pi$ and 12 $\sigma$ orbitals in an energy range from the Fermi energy to 10 eV binding energy have been used to benchmark four  exchange-correlations functionals commonly applied in density functional calculations. 
For the investigated bisanthene/Cu(110) system, the range-separated hybrid functional HSE was found to show the best performance regarding the energy level alignment of both the $\pi$ and $\sigma$ orbitals. 

While the present work has focused on the occupied molecular orbitals, first steps to extend POT to unoccupied states have already been undertaken. 
By transiently exciting electrons into unoccupied orbitals, the measurement of momentum signatures in excited states has recently been demonstrated by pump-probe angle-resolved photoemission experiments. \cite{Wallauer2020,Baumgartner2022,Neef2022} 
We envision that in future this will provide equally stringent experimental information for benchmarking the performance of  electronic structure methods describing optically excited states.

\acknowledgements
This work was funded by the Austrian Science Fund (FWF) project I3731 and the Deutsche Forschungsgemeinschaft (DFG) projects Po 2226/2-1, 223848855-SFB 1083 and  Ri 804/8-1.
The computations have been performed on the Vienna Scientific Computer (VSC) and the HPC facilities of the University of Graz. 
We thank  J. Riley (La Trobe University, Australia) for experimental support.


%

\end{document}